# Linear optics implementation of general two-photon projective measurement

Andrzej Grudka* and Antoni Wójcik**

Faculty of Physics, Adam Mickiewicz University,

Umultowska 85, 61-614 Poznań, Poland

Abstract

We will present a method of implementation of general projective measurement of two-photon polarization state with the use of linear optics elements only. The scheme presented succeeds with a probability of at least $1/16$. For some specific measurements, (e.g. parity measurement) this probability reaches $1/4$.

Recently, Hofmann and Tekeuchi [1] have presented a method of constructing the operator (the so called parity filter) projecting the input state of two photons on a two-dimensional subspace of identical horizontal and vertical polarizations. They have achieved the goal with the use of linear optics only. The necessary two-photon interactions have been simulated by detection and post selection process. The restriction to the linear optics is reflected in the probability of success of Hoffman and Takeuchi procedure which is equal to $1/16$. It is the aim of this paper to present a scheme which would allow to perform any projective two-photon polarization measurement with the same probability. Moreover in the specific case of the parity filter the scheme presented reaches the probability of success equal to $1/4$. The scheme is based on quantum teleportation in a way similar to the linear optics implementation of unitary quantum gates [2-7]. All these schemes use auxiliary photons. This photon redundancy allows performance of two-photon operations by appropriately chosen measurements. The result of the measurement outcome tells us which unitary operation has been performed.

In general the scheme presented consists of the following steps. First, one has to prepare a specific state of six auxiliary photons according to which kind of measurement is to be performed. It is worth noting, that any such state can be obtained non deterministically by the method of trial and error without nonlinear photon-photon interactions. Secondly, one performs two Bell measurements and two single-photon measurements.

Any operator which projects a state in a four-dimensional Hilbert space on some its subspace can be constructed as a sum of operators projecting this state on one-dimensional orthogonal subspaces. So, let us suppose that we have four orthonormal two-photon polarization states $|\alpha^i\rangle$ i.e. $\langle\alpha^i|\alpha^j\rangle = \delta_{ij}$. Then four operators that project on these states are $P^i = |\alpha^i\rangle\langle\alpha^i|$. The general form of the operator $\hat{P}^j$ which projects on subspace spanned by any subset of $\{|\alpha^i\rangle, i = 0,1,2,3\}$ is $\hat{P}^j = \sum_i \pi_i^j P^i$, where $\pi_i^j = 1$ if $|\alpha^i\rangle$ belongs to the subset defining $\hat{P}^j$ and $\pi_i^j = 0$ in the other case. In this way one can construct a family of projectors $\hat{P}^j$ which satisfy the condition $\sum_j \hat{P}^j = I$ provided that $\sum_j \pi_i^j = 1$. One can arrange projective measurement in which the initial state $|\beta\rangle$ is transformed to $\hat{P}^j|\beta\rangle = \sum_i \pi_i^j \langle\alpha^i|\beta\rangle|\alpha^i\rangle$ (up to normalization factor) with the probability $\langle\beta|\hat{P}^j|\beta\rangle$.

Let us now describe the first step of our scheme which we call the preparation step. Its aim is to prepare a specific auxiliary state of six photons (which will be labeled by indices 3,...,8) for a given family of projectors $\hat{P}^j$. Any orthonormal base for two-photon polarization space which defines projectors $P^i = |\alpha^i\rangle\langle\alpha^i|$ is of the form

$$|\alpha^i\rangle_{34} = \alpha^i_{00}|H\rangle_3|H\rangle_4 + \alpha^i_{01}|H\rangle_3|V\rangle_4 + \alpha^i_{10}|V\rangle_3|H\rangle_4 + \alpha^i_{11}|V\rangle_3|V\rangle_4, \quad (1)$$

where $i = 0,1,2,3$ and $H,V$ stand for horizontal and vertical polarization, respectively. We will also introduce another orthonormal base

$$|\alpha^i\rangle^*_{56} = (\alpha^i_{00})^*|V\rangle_5|V\rangle_6 + (\alpha^i_{01})^*|V\rangle_5|H\rangle_6 + (\alpha^i_{10})^*|H\rangle_5|V\rangle_6 + (\alpha^i_{11})^*|H\rangle_5|H\rangle_6 \quad (2)$$

and the following two-photon states

$$\begin{aligned}|0\rangle_{78} &= |H\rangle_7|H\rangle_8 \\ |1\rangle_{78} &= |H\rangle_7|V\rangle_8 \\ |2\rangle_{78} &= |V\rangle_7|H\rangle_8 \\ |3\rangle_{78} &= |V\rangle_7|V\rangle_8.\end{aligned} \quad (3)$$

The state which is needed to perform projective measurement is

$$|\Xi\rangle = \frac{1}{2}\sum_j\sum_i \pi^j_i |\alpha^i\rangle_{34}|\alpha^i\rangle^*_{56}|j\rangle_{78}. \quad (4)$$

As recently shown by several groups [3-6], it is possible to perform probabilistic CNOT polarization gate with the use of linear optics. It is well known [8] that CNOT gate together with one-photon gates (which can be easily performed) are universal. Thus, one can construct the required state by the method of trial and error. The point is that the preparation step does not involve the state which is to be measured. One first prepares the auxiliary state $|\Xi\rangle$, and then uses it. The probability of preparation of the state $|\Xi\rangle$ in one trial does not affect the probability of our scheme. We are now ready to perform the projective measurement on an arbitrary two-photon polarization state

$$|\beta\rangle_{12} = \beta_{00}^i |H\rangle_1 |H\rangle_2 + \beta_{01}^i |H\rangle_1 |V\rangle_2 + \beta_{10}^i |V\rangle_1 |H\rangle_2 + \beta_{11}^i |V\rangle_1 |V\rangle_2 .$$ (5)

To see how our scheme works we will write the total state of photons to be measured and auxiliary photons as

$$|\beta\rangle_{12} |\Xi\rangle_{345678} = \frac{1}{4} |\Psi^+\rangle_{15} |\Psi^+\rangle_{26} |\xi\rangle_{3478} + \text{the other terms} ,$$ (6)

where

$$|\Psi^\pm\rangle_{\mu\nu} = \frac{1}{\sqrt{2}} \left( |H\rangle_\mu |V\rangle_\nu \pm |V\rangle_\mu |H\rangle_\nu \right)$$ (7)

are Bell states, $|\xi\rangle_{3478}$ is the following four-photon normalized state

$$|\xi\rangle_{3478} = \sum_j \sum_i \pi_i^j \langle \alpha^i | \beta \rangle | \alpha^i \rangle_{34} | j \rangle_{78}$$ (8)

and *the other terms* contain only states orthogonal to $|\Psi^+\rangle_{15} |\Psi^+\rangle_{26}$. It is well known [9] that with the use of linear optics it is possible to identify two of four Bell states, namely $|\Psi^+\rangle$ and $|\Psi^-\rangle$. If one performs Bell measurements on photon pairs (1,5) and (2,6) and the measurements' results give $|\Psi^+\rangle$ for both pairs then the state of the remaining photons is projected on $|\xi\rangle_{3478}$. As can be seen from Eq. (6) this happens with the probability $1/16$. The state $|\xi\rangle_{3478}$ can be written in the form

$$|\xi\rangle_{3478} = \sum_j \left( \hat{P}^j |\beta\rangle_{34} \right) | j \rangle_{78} ,$$ (9)

which clearly indicates that the measurement performed on photons 7 and 8 in $\{|H\rangle, |V\rangle\}$ basis projects photons 3 and 4 on the state $\hat{P}^j |\beta\rangle$ provided that photons 7 and 8 are found in the state $|j\rangle$. The probability of projection on a given subspace is equal to $\langle \beta | \hat{P}^j | \beta \rangle$ in accordance with the standard formula. Thus, the aim of our scheme is achieved. The scheme can be straightforward generalized to $n$-photon state, however the probability of success scales as $4^{-n}$.

Let us now consider a specific case of measurement where the family of projectors consists of two elements only (the so called parity operators) given by the following formulae

$$\hat{P}^0 = |H\rangle|H\rangle\langle H|\langle H| + |V\rangle|V\rangle\langle V|\langle V|$$
$$\hat{P}^1 = |H\rangle|V\rangle\langle V|\langle H| + |V\rangle|H\rangle\langle H|\langle V|. \tag{10}$$

In this case $|\alpha^0\rangle_{34} = |H\rangle_3|H\rangle_4$, $|\alpha^1\rangle_{34} = |V\rangle_3|V\rangle_4$, $|\alpha^2\rangle_{34} = |H\rangle_3|V\rangle_4$, $|\alpha^3\rangle_{34} = |V\rangle_3|H\rangle_4$, $\pi_0^0 = \pi_1^0 = \pi_2^1 = \pi_3^1 = 1$, $\pi_0^1 = \pi_1^1 = \pi_2^0 = \pi_3^0 = 0$, and $j = 0,1$ so the states $|j\rangle$ can be encoded in one-photon polarization states $|j=0\rangle_7 = |H\rangle_7$, $|j=1\rangle_7 = |V\rangle_7$. The six-photon auxiliary state $|\Xi\rangle$ is now replaced by the five-photon one, namely

$$|\Xi\rangle_{34567} = \frac{1}{2}\left(|H\rangle_3|H\rangle_4|V\rangle_5|V\rangle_6 + |V\rangle_3|V\rangle_4|H\rangle_5|H\rangle_6\right)|H\rangle_7 + $$
$$+ \frac{1}{2}\left(|H\rangle_3|V\rangle_4|V\rangle_5|H\rangle_6 + |V\rangle_3|H\rangle_4|H\rangle_5|V\rangle_6\right)|V\rangle_7 \tag{11}$$

The total state of two photons to be measured and auxiliary photons can be written as

$$|\beta\rangle_{12}|\Xi\rangle_{34567} = \frac{1}{4}|\Psi^+\rangle_{15}|\Psi^+\rangle_{26}|\xi^0\rangle_{347} + $$
$$+ \frac{1}{4}|\Psi^+\rangle_{15}|\Psi^-\rangle_{26}|\xi^1\rangle_{347} + $$
$$+ \frac{1}{4}|\Psi^-\rangle_{15}|\Psi^+\rangle_{26}|\xi^2\rangle_{347} + \tag{12}$$
$$+ \frac{1}{4}|\Psi^-\rangle_{15}|\Psi^-\rangle_{26}|\xi^3\rangle_{347} + $$
$$+ \text{the other terms}$$

where

$$|\xi^0\rangle_{347} = \left(\beta_{00}|H\rangle_3|H\rangle_4 + \beta_{11}|V\rangle_3|V\rangle_4\right)|H\rangle_7 + $$
$$+ \left(\beta_{01}|H\rangle_3|V\rangle_4 + \beta_{10}|V\rangle_3|H\rangle_4\right)|V\rangle_7$$

$$|\xi^1\rangle_{347} = (\beta_{00}|H\rangle_3|H\rangle_4 - \beta_{11}|V\rangle_3|V\rangle_4)|H\rangle_7 +$$
$$+ (-\beta_{01}|H\rangle_3|V\rangle_4 + \beta_{10}|V\rangle_3|H\rangle_4)|V\rangle_7$$

$$|\xi^2\rangle_{347} = (\beta_{00}|H\rangle_3|H\rangle_4 - \beta_{11}|V\rangle_3|V\rangle_4)|H\rangle_7 +$$
$$+ (\beta_{01}|H\rangle_3|V\rangle_4 - \beta_{10}|V\rangle_3|H\rangle_4)|V\rangle_7 \quad (13)$$

$$|\xi^3\rangle_{347} = (\beta_{00}|H\rangle_3|H\rangle_4 + \beta_{11}|V\rangle_3|V\rangle_4)|H\rangle_7 +$$
$$+ (-\beta_{01}|H\rangle_3|V\rangle_4 - \beta_{10}|V\rangle_3|H\rangle_4)|V\rangle_7$$

while *the other terms* contain only states orthogonal to $|\Psi^\pm\rangle_{15}|\Psi^\pm\rangle_{26}$. In Eq. 12, analogous to Eq. 6, we have included $|\Psi^-\rangle$ terms for the reasons that become apparent later. Bell measurement on photon pairs (1,5) and (2,6) projects the remaining photons on $|\xi^0\rangle_{347}$ provided that the measurement outcome is $|\Psi^+\rangle_{15}|\Psi^+\rangle_{26}$. Moreover, if the result of Bell measurement is $|\Psi^+\rangle_{15}|\Psi^-\rangle_{26}$, $|\Psi^-\rangle_{15}|\Psi^+\rangle_{26}$ or $|\Psi^-\rangle_{15}|\Psi^-\rangle_{26}$ the remaining photons are in the state $|\xi^1\rangle_{347}$, $|\xi^2\rangle_{347}$ or $|\xi^3\rangle_{347}$, respectively, which can be easily transformed to the state $|\xi^0\rangle_{347}$ by the following one-photon unitary operation

$$Z_4|\xi^1\rangle_{347} = |\xi^0\rangle_{347}$$
$$Z_3|\xi^2\rangle_{347} = |\xi^0\rangle_{347} \quad (14)$$
$$Z_3 Z_4|\xi^3\rangle_{347} = |\xi^0\rangle_{347}$$

In these equations $Z_\mu$ acts only on the $\mu$-th photon and

$$Z_\mu|H\rangle_\mu = |H\rangle_\mu$$
$$Z_\mu|V\rangle_\mu = -|V\rangle_\mu. \quad (15)$$

The measurement performed on photon 7 completes the scheme, which in this particular case succeeds with the probability $1/4$.

Finally we would like to mention that if we restrict ourselves to perform the parity filter operation $\hat{P}^0 = |H\rangle|H\rangle\langle H|\langle H| + |V\rangle|V\rangle\langle V|\langle V|$ as in the case of Hofmann's and Takeuchi's scheme, we can further simplify the auxiliary photon state as follows

$$|\Xi\rangle_{3457} = \frac{1}{\sqrt{2}}\left(|H\rangle_3|H\rangle_4|V\rangle_5|V\rangle_6 + |V\rangle_3|V\rangle_4|H\rangle_5|H\rangle_6\right). \tag{16}$$

This four-photon entangled state is the cost we have to pay (in comparison with two auxiliary photons used in the scheme of Hofmann and Takeuchi) in order to perform parity filter operation with the probability $1/4$ instead of Hofmann's and Takeuchi's $1/16$.

In conclusion, we have presented linear optics implementation of general projective measurement of two-photon polarization state, which succeeds with a probability of at least $1/16$. We have also shown that in the case of the realization of the parity operator the probability of success is $1/4$.

We would like to thank the Polish Committee for Scientific Research for financial support under grant no. 0 T00A 003 23.

*Email address: agie@amu.edu.pl

**Email address: antwoj@amu.edu.pl